\documentclass[twocolumn,superscriptaddress]{revtex4-2}

\usepackage{svg}

\usepackage{amsmath,braket,dsfont,verbatim,latexsym,amssymb,indentfirst,mathrsfs,mathtools,amsthm,bbm,bm,hyperref,url,cancel,subcaption,enumitem,setspace}
\usepackage[font=small,labelfont=bf,format=plain,singlelinecheck=false,justification=centerlast]{caption}
\usepackage{verbatim,indentfirst}
\usepackage[title,titletoc]{appendix}
\usepackage{graphicx}
\usepackage{epstopdf}
\usepackage{float}
\usepackage{booktabs}
\usepackage{xr}
\usepackage{color}
\usepackage[dvipsnames]{xcolor}
\usepackage{adjustbox}

\begin{document}

\title{Time-Delocalized Local Measurements in an Indefinite Causal Order}

\author{Yann Valibouse}
\email{yann.valibouse@univie.ac.at}
\affiliation{University of Vienna, Faculty of Physics, Vienna Center for Quantum Science and Technology (VCQ), Boltzmanngasse 5, 1090 Vienna, Austria}
\affiliation{University of Vienna, Vienna Doctoral School in Physics, Boltzmanngasse 5, 1090 Vienna, Austria}

\author{Martí Cladera-Rosselló}
\affiliation{University of Vienna, Faculty of Physics, Vienna Center for Quantum Science and Technology (VCQ), Boltzmanngasse 5, 1090 Vienna, Austria}
\affiliation{University of Vienna, Vienna Doctoral School in Physics, Boltzmanngasse 5, 1090 Vienna, Austria}

\author{Michael Antesberger}
\affiliation{University of Vienna, Faculty of Physics, Vienna Center for Quantum Science and Technology (VCQ), Boltzmanngasse 5, 1090 Vienna, Austria}
\affiliation{University of Vienna, Vienna Doctoral School in Physics, Boltzmanngasse 5, 1090 Vienna, Austria}

\author{Patrick Lima}
\affiliation{University of Vienna, Faculty of Physics, Vienna Center for Quantum Science and Technology (VCQ), Boltzmanngasse 5, 1090 Vienna, Austria}
\affiliation{Departamento de F'isica, Universidade Federal de Minas Gerais, 31270-901, Belo Horizonte, Minas Gerais, Brazil}

\author{Philip Walther}
\affiliation{University of Vienna, Faculty of Physics, Vienna Center for Quantum Science and Technology (VCQ), Boltzmanngasse 5, 1090 Vienna, Austria}
\affiliation{Christian Doppler Laboratory for Photonic Quantum Computer, Boltzmanngasse 5, 1090 Vienna, Austria}
\affiliation{Institute for Quantum Optics and Quantum Information (IQOQI) Vienna, Austrian Academy of Sciences, Boltzmanngasse 3, 1090 Vienna, Austria}

\author{Lee A. Rozema}
\affiliation{University of Vienna, Faculty of Physics, Vienna Center for Quantum Science and Technology (VCQ), Boltzmanngasse 5, 1090 Vienna, Austria}

\date{\today}

\begin{abstract}
Processes with indefinite causal order (ICO), such as the quantum switch, are an emerging resource for quantum tasks and a fundamental test bed for studies of temporal correlations in quantum mechanics. 
A limitation of past photonic implementations of the quantum switch, however, is their inability to perform measurements inside the switch without either destroying the superposition of causal orders or delaying readout until the after the quantum switch.
Measurements where the results are read out locally are needed for several applications of ICO, but also for a loophole-free verification of ICO.
Here, we overcome past limitations by introducing a \textit{local} measurement scheme and coupling the photon in the switch to a \textit{time-delocalized} ancilla system.
We experimentally realize this protocol using a photonic quantum switch with post-selected linear optical logic gates.
Our method ensures that the measurement apparatus interacts with the system at two distinct times and yet yields a single outcome. 
We use a quantum eraser measurement to preserve the ICO, which we certify by measuring a causal witness and finding a negative value of $\mathcal{C}_W \approx -0.305 (1)$.
Furthermore, by explicitly realizing a time-delocalized ancilla system, our protocol not only enables a new class of quantum switch protocols requiring local readout, but also provides a general method for path-coherence-preserving measurements with broad applications beyond ICO.

\end{abstract}

\maketitle

Quantum mechanics allows for processes to occur in a indefinite causal order (ICO), which means processes can occur in a superposition of different orders \cite{chiribella2013quantum}.
ICO has attracted considerable attention as both a foundational concept \cite{oreshkov2012quantum} and a potential operational resource \cite{rozema2024experimental}.
From the foundational side, ICO is exciting as processes with an ICO generate a type of temporal correlations that are distinct from standard causally-ordered processes \cite{oreshkov2012quantum,baumeler2014maximal,baumeler2016space,wechs2021quantum}.
Although different processes with ICO have been proposed, most work has focused on a specific process called the quantum switch \cite{chiribella2013quantum}.
The quantum switch applies two operations to a target system in an order that depends on the state of a control qubit.
Since the quantum switch provides experimental access to a new quantum resource, there have been many proposals showing that the quantum switch can outperform causally-ordered processes at a variety of tasks \cite{bavaresco2022UnitaryChannelDiscrimination,Araujo2014,taddei2021computational,renner2022advantage,baumeler13,feixquantum2015,Guerin2016,Felce2021IBMswitch,Guha2020,Simonov2022WorkExtraction,Frey2019depolarizingChannelIdentification,ChapeauBlondeau2022,Zhao2020,spencer2025indefinite,Ning2023MeasuringIncompatibility}.
These proposals span a wide range, from computational and communication tasks to thermodynamics and quantum metrology.
This wide applicability, however, poses a major experimental challenge: although photonic quantum switches have been experimentally demonstrated \cite{lorenzo2015,rubino2017ExperimentalVerification,Rubino2021Communication,Rubino2022experimentalEntanglement,guo2020experimental,cao2022quantumSimulation,cao2022Semideviceindependent,zhu2023prl,Min23,wei2019experimentalCommunication,goswami2018Indefinite,goswami2020IncreasingCommunication,yin2023experimental,Antesberger2023tomography,richter2025towards}, different protocols require the superposition of vastly different channels.
For example, the first experiments demonstrated the superposition of unitary polarization channels \cite{lorenzo2015}, but experimental work has since progressed to more complex channels, including noisy channels \cite{Rubino2021Communication}, thermal reservoirs \cite{cao2022quantumSimulation}, and continuous variable phase-space displacements \cite{yin2023experimental}.

One class of channels of particular interest are the measurement and re-prepare channels, wherein the superposed party measures the target system and reads out the results.
If one wishes to view the superposed channels as parties or agents, rather than quantum gates, such channels are essential.
For example, causal inequalities (which are Bell-like inequalities for testing ICO) rely on these channels.
Although the quantum switch cannot violate causal inequalities \cite{araujo2015witnessing,Purves2021CannotViolate}, a device-independent verification of ICO with the quantum switch is possible, but it also requires superposed measurement and re-prepare channels \cite{van2023device}.
On top of the fundamental need to superpose measurement channels, they are also required for applications including characterizing ICO via higher-order process tomography \cite{Antesberger2023tomography}, measuring more robust causal witnesses \cite{araujo2015witnessing,bavaresco19}, indefinite causal key distribution \cite{spencer2025indefinite, lesniak2026bipartite}, and measuring the incompatibility of quantum observables \cite{Ning2023MeasuringIncompatibility}.

Although measurements have already been implemented in the quantum switch \cite{rubino2017ExperimentalVerification,cao2022Semideviceindependent,Antesberger2023tomography} all of these experiments suffer from the same drawback which prevents them from being used for a loophole-free demonstration of ICO. 
Namely, the measurement outcomes are accessed only after the quantum switch, and not \textit{locally} in the channel.
This is because the measurements implemented to-date couple the system qubit encoded in, for example, the polarization of a photon to a different internal degree of freedom of the \textit{same photon}.
Since only a single photon is used, the measurement result clearly cannot be read out until after the photon exits the quantum switch.
Moreover, one must ensure that measurements are performed without destroying path coherence. 
This leaves interpretational loopholes and limits the range of practically implementable protocols.
For example, indefinite causal key distribution places Alice and Bob in the quantum switch \cite{spencer2025indefinite}. 
If Bob cannot read out his measurement result locally any implementation will not be practical.  
On the other hand, in order to close all of the loopholes in a device-independent verification of ICO using the protocol of \cite{van2023device}, the parties inside of the quantum switch must read out their measurement results before the photon exits the switch \cite{richter2025towards}.

In this work, we overcome this limitation by showing how to implement a local measurement inside the quantum switch with a \textit{time-delocalized} ancillary system that stays with one party and interacts with the target system via two quantum logic gates, while still preserving coherence between the two causal orders. 
We then experimentally implement this proposal using a second photon, post-selected linear optical quantum logic gates, and a quantum eraser measurement to maintain the ICO.
We certify the ICO by evaluating a \emph{causal witness} to quantify the causal non-separability of the underlying process \cite{araujo2015witnessing,rubino2017ExperimentalVerification}.
We further show that removing the quantum eraser measurement increases the which-path information \cite{englert1996fringe}, driving a transition to classical mixture of causal orders. Our experiment illustrates the direct connection between coherence in the control degree of freedom and the ICO, showing how to use a time-delocalized system to maintain the ICO in a photonic quantum switch.

\begin{figure}
    \centering
    \includegraphics[width=.95\linewidth]{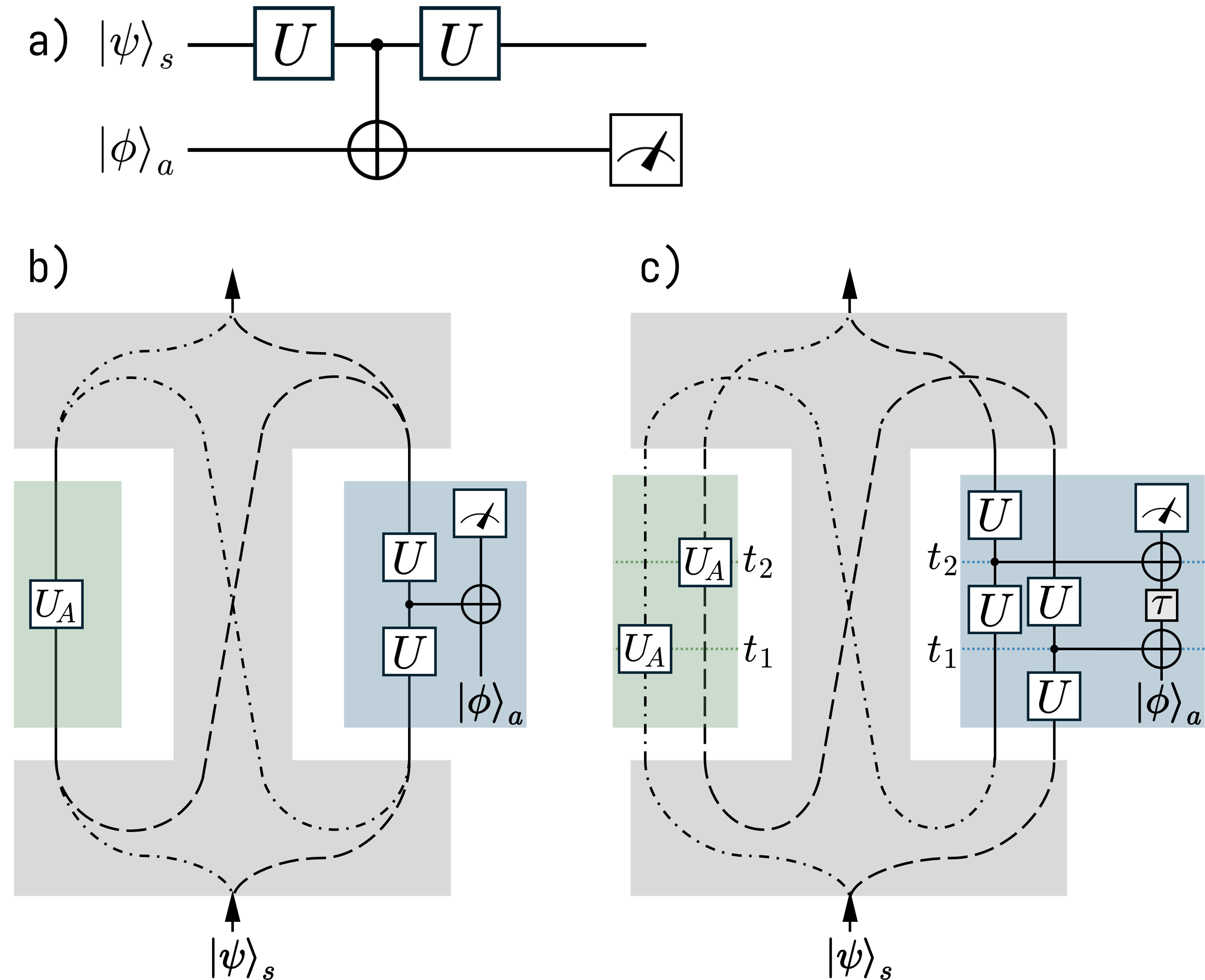}
    \caption{\textbf{Time-delocalized local von Neumann measurement in a quantum switch}. 
    \textit{(a)~von Neumann measurement:} a controlled-NOT gate entangles the system with a probe ancilla, whose readout collapses the system. The unitary gates $U$ are used to couple different observables of the system qubit to the ancilla. 
    \textit{(b)~Implementation within the quantum switch.} Alice (green) applies a unitary operation $U_A$, while Bob (blue) performs the von Neumann measurement described in (a). The quantum switch coherently controls the order of these operations, placing Alice's unitary and Bob's measurement in a superposition of the two possible causal orders.
    \textit{(c)~Time-delocalized implementation in the quantum switch.} In practice, the process unfolds at two times, $t_1$ and $t_2$. At time $t_1$, the system interacts in a coherent superposition of Alice and Bob (Alice in one branch and Bob in the other). At time $t_2$, the roles are exchanged, so that the system interacts with Bob in superposition with Alice. In this way, each operation is applied once, but its temporal location is delocalized across the two times $t_1$ and $t_2$. Bob's probe ancilla therefore couples to the system at $t_1$ in one branch and at $t_2$ in the other. A temporal delay $\tau$ is introduced to erase which-order information carried by the probe, restoring coherence between the two processes. The resulting measurement is a single effective von Neumann measurement, delocalized in time while remaining locally encoded in Bob's laboratory.}
    \label{fig:concept}
\end{figure}

\section{Time-Delocalized von Neumann Measurements}

We will start by briefly reviewing the von Neumann formalism, using quantum logic gates. In this scenario, we imagine measuring a system qubit $\ket{\psi}_s$ by coupling it to a probe ancilla qubit $\ket{\phi}_a$ via a controlled-not (CNOT) gate, as shown in Fig.~\ref{fig:concept}a.
Consider $\ket{\psi}_s = \alpha\ket{0}_s + \beta\ket{1}_s$, 
the ancilla qubit is initialized in the state $\ket{0}_a$, and $U=I$. 
This effectively couples the spin along $Z$ of the system to the ancilla qubit via a CNOT gate, which takes the place of the \textit{measurement interaction} introduced by von Neumann. 
In particular, this interaction produces the entangled state
$\alpha\ket{0}_s\ket{0}_a+\beta\ket{1}_s\ket{1}_a$. 
Measuring the expectation value of the Pauli $\sigma_z$ operator on the probe qubit yields 
$\langle\sigma_z\rangle = |\alpha|^2-|\beta|^2$,
which is exactly the result one would obtain by directly measuring the expectation value of $\sigma_z$ on the system qubit. Moreover, tracing out the ancilla leaves the system is in the mixed state
$|\alpha|^2\ket{0}\!\bra{0} + |\beta|^2\ket{1}\!\bra{1}$,
as required by the projection postulate, without physically measuring the system qubit.
If the gates $U$ in Fig. \ref{fig:concept}a are set to different operators, one can perfectly reconstruct different single-qubit expectation values of the system, without needing to physically measure the system.
In other words, this is a form of a quantum non-demolition measurement.
Now consider inserting this scheme inside the quantum switch, as depicted in Fig.~\ref{fig:concept}b. The quantum switch coherently controls the order in which Alice's unitary $U_A$ (green) and Bob's von Neumann measurement (blue) act on the system, placing them in superposition of the two causal orders.

In practice, this is realized through time-delocalized protocol shown in Fig.~\ref{fig:concept}c. The process develops across two times, $t_1$ then $t_2$. A photon encoding a qubit polarization degree of freedom (where $0$ is encoded in the horizontal polarization and $1$ in the vertical polarization) enters from the bottom. In Bob's lab a second photon is used to encode the ancilla qubit in its polarization.
We further imagine that Bob has perfect CNOT gates to implement the measurement interaction.

We will first work out the simple case where Alice performs the identity on the system state ($U_A=I$).
Prior to Bob's measurement, the joint state is
\begin{equation}
    \ket{\Psi} = \frac{1}{\sqrt{2}}(\ket{0}_c+\ket{1}_c) \otimes\ket{\psi}_s\otimes\ket{0}_a, 
\end{equation}
where $\ket{0}_c/\ket{1}_c$ denotes the control qubit encoded in the path of the first photon, and qubits $s$ and $a$ are the system and ancilla photons encoded in the polarization of photons 1 and 2, respectively.

Now at a first time step, $t_1$, one component of the system photon first passes through Bob's von Neumann measurement. The entangling interaction correlates the probe polarization with the system polarization in the causal order set by $\ket{1}_c$, leading to the state
\begin{align}
    \ket{\Psi}_{t_1} = \frac{1}{\sqrt{2}} \bigg( &\ket{0}_c(\alpha\ket{0}_s+\beta\ket{1}_s)\ket{0}_a \notag\\
    &+ \ket{1}_c(\alpha\ket{0}_s\ket{0}_{a}+\beta\ket{1}_s\ket{1}_{a})\bigg) 
\end{align}
Notice that at this point, there is which-path information encoded in the ancilla qubit, i.e. tracing out the ancilla and system qubits would, in general, leave the control qubit in a mixed state.

Now at the second time step, $t_2$, the second component of the system arrives at Bob's station.
Here it is essential that the ancilla qubit is delayed (denoted by the box labelled $\tau$ in Fig.~\ref{fig:concept}c), such that it arrives at Bob's second CNOT gate at the same time as the photon encoding the system qubit.
If this is the case, then the which-path information is erased:
\begin{equation}
    \ket{\Psi}_{t_2} = \frac{1}{\sqrt{2}} (\ket{0}_c+\ket{1}_c)(\alpha\ket{0}_s\ket{0}_{a}+\beta\ket{1}_s\ket{1}_{a});
\end{equation}
i.e. this state exhibits coherence in the control qubit encoded in the path of the photon.

This two-step measurement procedure constitutes a \emph{time-delocalized measurement}. The system photon is measured exactly once (yielding a single value encoded in the probe), but the measurement is not associated with a single interaction event at a definite time. Instead, it is realized through two probe interactions (at $t_1$ and $t_2$) that together constitute a single measurement operation delocalized across two times.
This time delocalization is a structural feature of photonic implementations of the quantum switch. 
As shown by Oreshkov \cite{Oreshkov2019time}, the input and output systems of a party inside the quantum switch are, in general, time-delocalized subsystems whose causal structure can be described by the process matrix framework \cite{oreshkov2012quantum}. 
Probing the ancilla between $t_1$ and $t_2$ would correspond to accessing the system's degrees of freedom before the delocalized operation is complete, leaking which-path information and degrading the ICO.
At the same time, this measurement scheme is local, in the sense that the ancilla photon can physically remain in Bob's measurement station.

This measurement scheme also works in the case where Alice performs any operation.  For example, if she implements a unitary $\hat{U}_A$ on the system polarization it is straightforward to see that the resulting state at $t_2$ is
\small
\begin{align}
    \ket{\Psi}_{t_2} = \frac{1}{\sqrt{2}} \bigg( &\ket{0}_c\text{CNOT}(\hat{U}_A\otimes\mathbb{I})\ket{\varphi}_s\ket{0}_a \notag\\
    &+ \ket{1}_c(\hat{U}_A\otimes\mathbb{I})\text{CNOT}\ket{\varphi}_s\ket{0}_a \bigg) 
\end{align}
\normalsize
which can be factorized only if $\hat{A}$ and the CNOT commute or anticommute, as expected from the usual commutation games inside a quantum switch \cite{lorenzo2015}.
We show below that this is sufficient to violate a causal witness.

\section{Linear Optical Implementation}

\begin{figure}
    \centering
    \includegraphics[width=\linewidth]{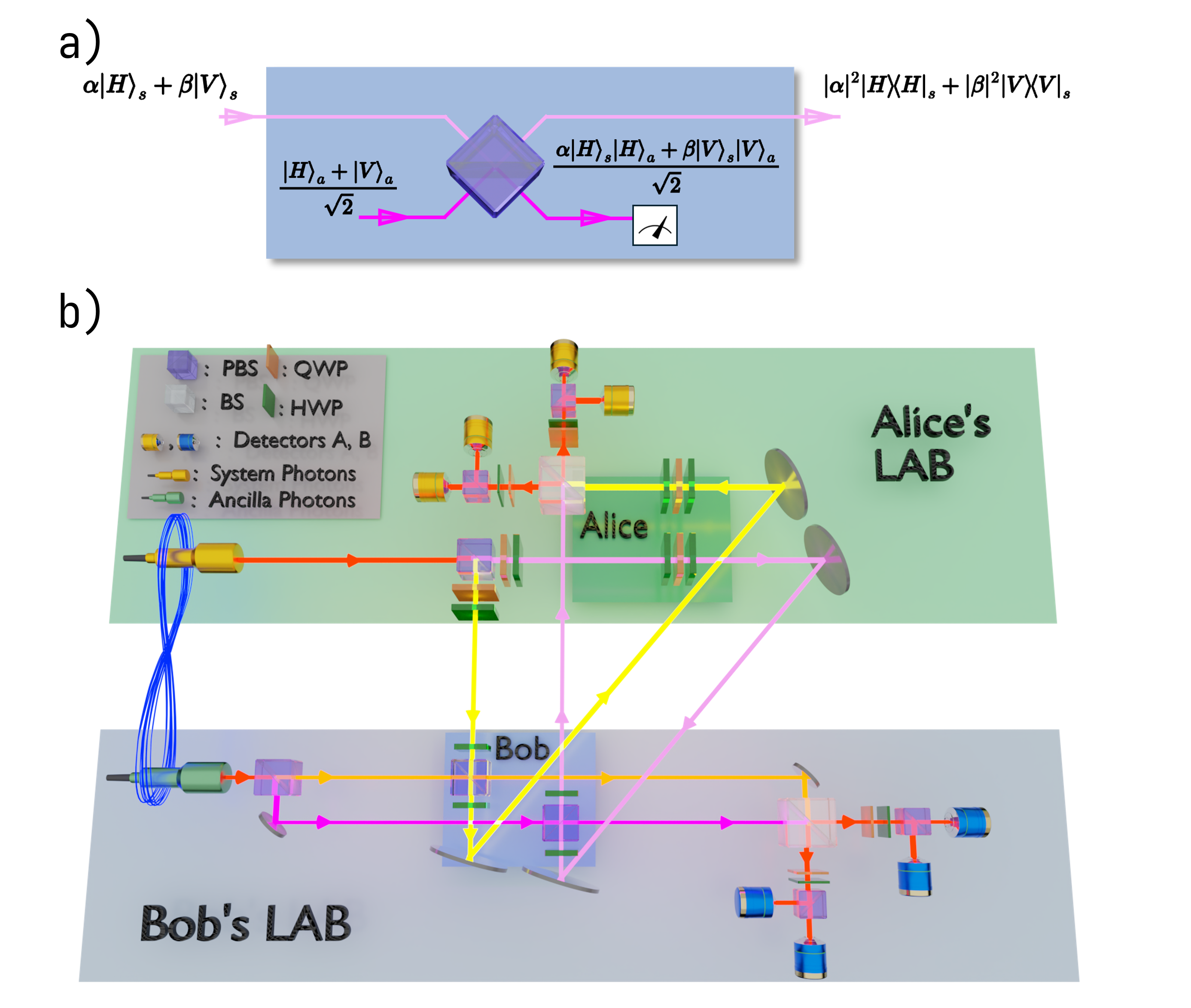}
    \caption{\textbf{Experimental implementation}. 
    \textit{(a)~Non-destructive photonic implementation of a von Neumann measurement of a single-photon polarization.} The system photon is coupled to an ancilla via a post-selected CNOT-like gate implemented with a polarizing beam splitter, whose measurement basis can be tuned by waveplates. The polarization information is retrieved by measuring the ancilla, which correspondingly projects the system state.
    \textit{(b)~Post-selected von Neumann measurement within the quantum switch.} We exploit path entanglement as a resource to enable successful post-selection, allowing two successive non-deterministic CNOT gates, as in (a), to implement a local measurement of the system polarization inside the switch. The system and ancilla photons are entangled in their path degree of freedom (as indicated by the path colors). After the measurement interaction at Bob’s polarizing beam splitter, the ancilla paths are recombined to erase which-path information and restore coherence. Alice’s unitaries are implemented using a quarter--half--quarter waveplate sequence.
    }
    \label{fig:setup_}
\end{figure}

Experimentally, it is challenging, although not fundamentally impossible, to implement deterministic two-photon logic gates. We therefore rely on post-selected linear-optical gates \cite{KLM,ralph2002linear}, which have been used to implement non-destructive von Neumann measurements of single-photons \cite{pryde2004measuring,pryde2005measurement}. 
While the standard linear-optical CNOT gate succeeds with probability $1/9$, our protocol does not require a full CNOT gate if the state of the ancilla qubit is fixed \cite{pittman2001probabilistic}. 
As a result, the entangling interaction can be implemented with a simpler post-selected scheme whose success probability is $1/2$.

In our experiment, we fix the state of the ancilla photon to
$\ket{D}_a = \tfrac{1}{\sqrt{2}}(\ket{H}_a + \ket{V}_a)$, while the system photon can be in an arbitrary pure state
$\ket{\psi}_s = \alpha \ket{H}_s + \beta \ket{V}_s$. By sending these photons into the two input ports of a PBS (Fig.~\ref{fig:setup_}a), and post-selecting on events where there is one photon per output port (success probability of $1/2$), we entangle the polarization of two photons: 
$\ket{\psi'} = \alpha \ket{H}_s \ket{H}_a + \beta \ket{V}_s \ket{V}_a$, just as in the von Neumann measurement using a CNOT gate discussed above.
To perform a change of basis, waveplates can be placed before and after the PBS on the system paths. 
These two waveplates can be set to different angles, allowing one to implement a measurement in a certain basis (defined by the waveplate before the PBS) then reprepare the system in another basis (defined by the waveplate after the PBS).
Note that this entanglement generation requires two-photon interference at the PBS: the post-selection only projects the separable two photon state on the polarization-entangled state
when the photons are indistinguishable and arrive simultaneously.

In the ideal protocol, two consecutive logic gates are used with the ancilla qubit as the target.
With post-selected logic gates this is not possible, as it prevents one from determining if the first gate was successful since the failure modes can mix after the photons propagate through the second gate \cite{barz2014two}. 
To circumvent this problem, we prepare the probe photon in a superposition of paths that must be correlated with the paths of the system photon in the switch. 
In other words, as shown in Fig.~\ref{fig:setup_}b, we entangle the path degrees of freedom the system and ancilla photons such that when the system photon is the yellow (pink) paths Bob's ancilla photon is the orange (magenta) paths and the photons will meet at the correct PBS in Bob's lab.

Experimentally, we achieve this by generating polarization entangled photons with a Type-II SPDC source based on a 3cm long PPKTP crystal in a Sagnac geometry, pumped at 775nm to produce photon pairs at 1550nm (not pictured).
Then each photon is sent to a PBS, after which their polarization is erased with a waveplate in the reflected path so that only the path remains entangled, see Fig.~\ref{fig:setup_}b. 

For photon 1 in the quantum switch, we then use quarter and half waveplates in each output port of the first switch PBS to set arbitrary system states. Thus, the general state before the measurement interaction is
\begin{equation}
    \frac{1}{\sqrt2}(\ket{\psi,0}_{s,c} \otimes \ket{D,0}_{a,p} + \ket{\psi,1}_{s,c} \otimes \ket{D,1}_{a,p}).
\end{equation}
Here $s$ is the system qubit (encoded in the first photon's polarization), $c$ is the control qubit of the quantum switch (encoded in the path of the first photon), $a$ is the state of the ancilla qubit (encoded in the polarization of the second photon), and $p$ is the path of the second photon (which is entangled with the control qubit).

The path entanglement is more apparent when factored according to path and polarization degrees of freedom:
\begin{equation}
    \ket{\psi}_{s}\ket{D}_a (\frac{\ket{0,0}_{c,p} + \ket{1,1}_{c,p}}{\sqrt2}),
\end{equation}

After the measurement interaction, the polarization of the ancilla becomes entangled with the polarization of the system photon in their respective paths, and the two-photon state reads,
\begin{equation}
    \frac{1}{\sqrt{2}}\big(\ket{\psi_{A\rightarrow B}}_{s,a}\ket{0,0}_{c,p} + \ket{\psi_{B\rightarrow A}}_{s,a}\ket{1,1}_{c,p}\big)
\end{equation}
where $\ket{\psi_{A\rightarrow B}}_{s,a}$ and $\ket{\psi_{B\rightarrow A}}_{s,a}$ are two polarization entangled states, corresponding to Alice acting before or after Bob interacts his ancilla qubit with the system photon, respectively.
The full states are fully written in the Appendix Eq.~\ref{bob_state}.

Since the ancilla photon's path remains entangled with the control qubit it stores which path information. 
In practice, this means that tracing over it would destroy the ICO.
To remedy this, we erase the which-path information by interfering the two paths of the ancilla photon on a second beamsplitter in Bob's laboratory. 
In the ideal proposal (Fig. \ref{fig:concept}a), this erasure is done deterministically by setting $\tau$ appropriately.
In our experimental implementation, the output ports of the beamsplitter at the end of the quantum switch become correlated with the output ports of Bob's ancilla beamsplitter.

Nevertheless, measurements of the ancilla do not reveal any which-path information about the system photon and therefore preserve the ICO of the switch. 

After the measurement procedure, post-selection, and path erasure, we can measure the polarization of the photon inside the quantum switch via the ancilla without destroying the ICO. See the Appendix for details. We point out that our measurement scheme is quite general, and thus can be applied to photons within various interferometer geometries, while ensuring the path coherence is preserved.

\begin{figure}[t]
    \centering
    \adjustbox{trim=2.28cm 0.3cm 0cm 0.76cm, clip, width=\linewidth}{
        \includegraphics[width=\linewidth]{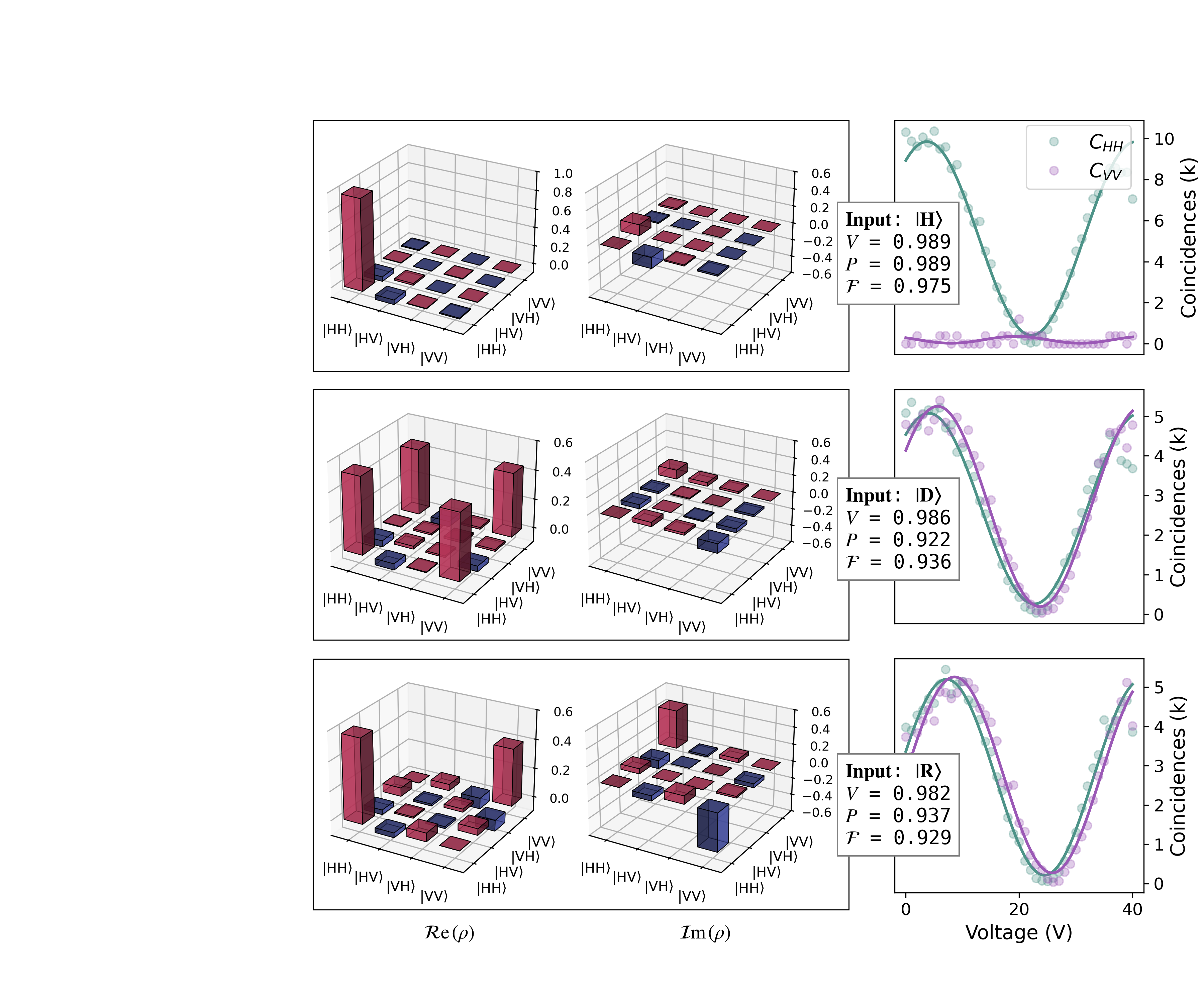}    
    }
    
    \caption{\textbf{Characterization of Bob's Measurements.} \textit{Left plots}: Two-photon tomography between Bob's ancilla and Alice system after the switch for three different inputs.
    \textit{Right plots}: Interference fringes of the quantum switch in the presence of Bob's measurements. The phase in the interferometer is scanned via a voltage-controlled piezzo in one arm of the the quantum switch. 
    We display the coincidence counts between the two transmitted ports $C_{HH}$ of the tomography setups and the two reflected ports $C_{VV}$.
    $\boldsymbol{V}$ is the visibility of the interference,  $\boldsymbol{P}$ the purity of the two-photon state, 
    $\boldsymbol{\mathcal{F}}$ is the fidelity to expected states, described in the main text.}
    \label{fig:tomo_scan}
\end{figure}

\section{Results}

As a first step to verify that Bob's measurement scheme is faithful, we fix Bob to measure and reprepare in the computational basis, we have Alice implement the identity operation, and we prepare the system photon at the input of the quantum switch in one of the states $\ket{H}_s$, $\ket{D}_s$, or $\ket{R}_s$.
We then perform two-photon quantum state tomography on system qubit (the first photon's polarization) and Bob’s ancilla photon at the output ports of their respective beam splitters. Due to the entangling mechanism described above, initialization in $\ket{H}_s$ is expected to yield the separable state $\ket{H}_s\ket{H}_a$, whereas initialization in $\ket{D}_s$ ($\ket{R}_s$)  should result in the maximally entangled state
$\tfrac{1}{\sqrt{2}}(\ket{H}_s\ket{H}_a + \ket{V}_s\ket{V}_a)
\quad
\text{( }
\tfrac{1}{\sqrt{2}}(\ket{H}_s\ket{H}_a - i\ket{V}_s\ket{V}_a)\text{)}
$.
The results of these quantum state tomography measurements are presented in left column of Fig. \ref{fig:tomo_scan}.
We observe two-photon state fidelities and purities above $0.92$ for all input states (see inset).
The reduction in purity and fidelity of the entangled states primarily comes from phase fluctuations of the interferometers which remain passively stable for more than 30 minutes.

To also verify that these measurements do not reveal which-path information, we perform a fringe scan by varying the phase in the quantum switch. This is achieved by translating a mirror mounted on a piezoelectric actuator. We do this for each input polarization, as Bob simultaneously measures the switch photon's polarization. These scans reveal interference visibilities above $0.98$ (Fig.~\ref{fig:tomo_scan}, right). These measurements confirm that Bob can measure the switch photon's polarization locally without destroying the path coherence and thus suggest that our quantum switch possesses an ICO.

Next, to rigorously prove the ICO, we measure a \emph{causal witness} $S$ (see Appendix) \cite{araujo2015witnessing,rubino2017ExperimentalVerification}. 
Within the process matrix framework, a process matrix $W$ is used to describe the quantum switch.
Then, in analogy with entanglement witnesses for quantum states, a negative expectation value of the causal witness $\mathcal{C}_W=\mathrm{Tr}(S W)$ certifies the \textit{causal non-separability} of a process matrix (a measure of the ICO), while if $\mathcal{C}_W \ge 0$ the experiment could be in a classical mixture of definite causal orders.

To measure $\mathcal{C}_W$, Alice and Bob must implement different operations.
Additionally, the input state and measurements of the control qubit after the switch can be varied.
The precise value of $\mathcal{C}_W$ depends on the available measurement settings.
In our experiment, Bob perform measurements in the $Z$ and $X$ polarization bases and can reprepare the system in eigenstates of either of these bases (implemented using half waveplates placed before and after the PBS). Alice implements arbitrary unitary operations using a quarter--half--quarter waveplate configuration, which we use to realize a set of ten unitaries described in the Appendix. Finally, the input state is prepared in $\ket{H}_s$, $\ket{D}_s$, or $\ket{R}_s$, while the control qubit is always measured in the $X$ basis. The result is $180$ different
experimental settings, each with four outcomes and thus four probabilities to estimate (Bob's has two outcomes and the control qubit measurement has two outcomes).

\begin{figure}[t]
    \centering
    \includegraphics[width=\linewidth]{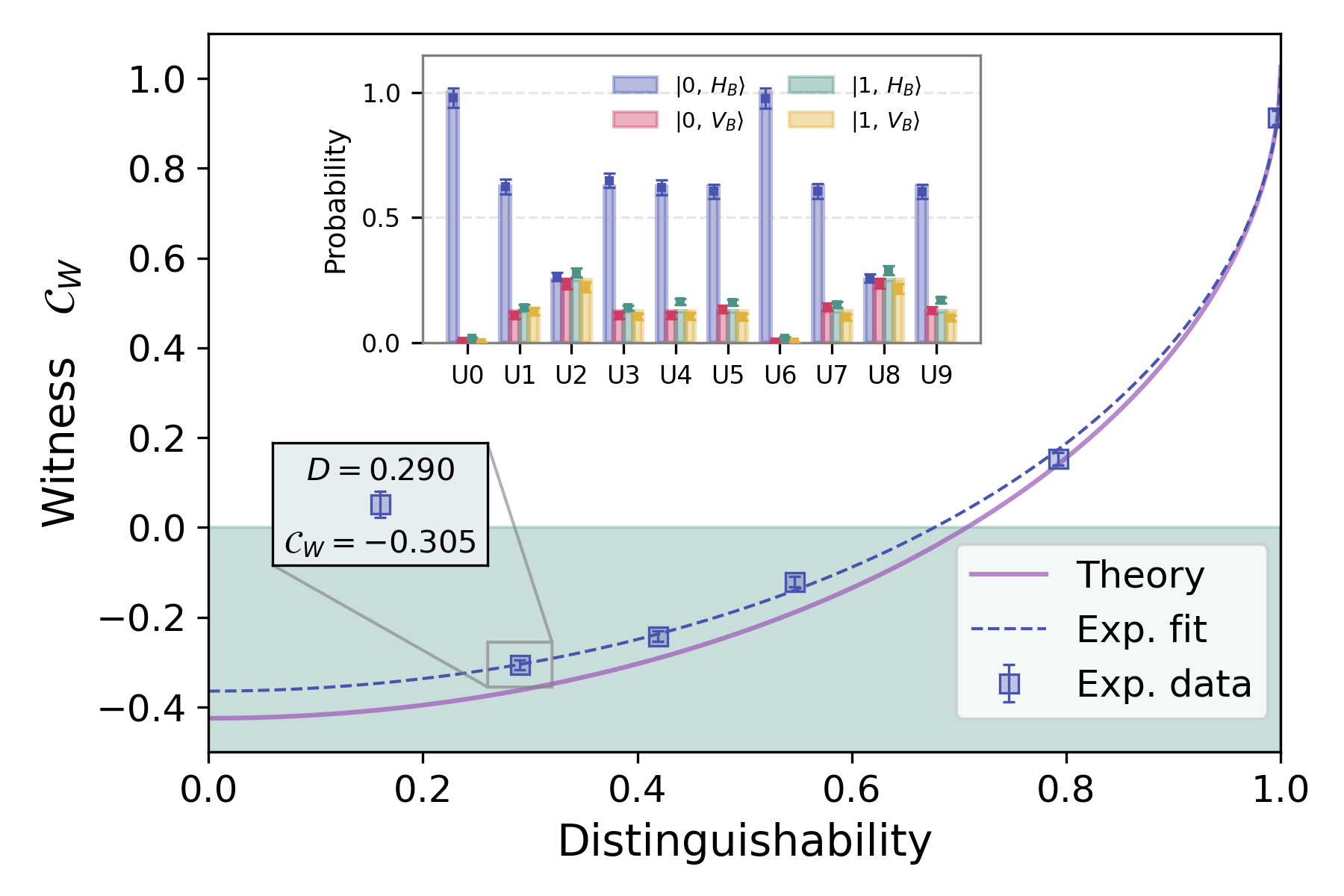}
    \caption{\textbf{Causal witness in the presence of which-path information}. For a quantum switch visibility of $0.957$, the distinguishability is $\approx 0.29$ for a pure state, leading to $\mathcal{C}_W=-0.305 (1)$, which is deep in the non-causal region indicated in mint blue.
    \textit{Inset:} The measured probabilities for the four possible outcomes for each of Alice's ten unitaries operation when Bob measures and reprepares in $Z$ and the input state is set to $\ket{H}$.
    The witness measurement is based on these data, together with the additional plots shown in the Appendix. }
    \label{fig:wit_vs_vis}
\end{figure}

From these settings, we use a semidefinite program to construct an optimal causal witness matrix $S$, as described in \cite{rubino2017ExperimentalVerification, Antesberger2023tomography}. 
In practice, this provides a set of coefficients used to weight our experimentally measured probabilities (See Appendix eq.~\ref{causal_witness}). 
For each experimental configuration, the measurement duration is 3 seconds with an average photon pair rate of $\approx 10,000$ pairs/s.
The inset of Fig. \ref{fig:wit_vs_vis} shows a representative set of these experimental probabilities, showing good agreement between theory (bars) and experiment (points).
We obtain an experimental value of the causal witness $\mathcal{C}_W^{(\mathrm{Exp})} \approx -0.305(1)$, while the minimal theoretical value is $\mathcal{C}_W^{(\mathrm{Th})} = -0.4248$. The discrepancy between the theoretical and experimental values arises mainly from the limited purity of the
input two-photon state and from imperfections in the interferometric implementation of the quantum switch and Bob’s measurement stage. In particular,
the interferometers do not exhibit perfect visibility, and their visibility is sensitive to interferometric phase fluctuations. These effects reduce the precision of the measured probabilities and consequently increase the observed value of the causal witness.\\

As we have already alluded to, in photonic implementations of the quantum switch the ICO is dependent on ensuring that both orders are superposed without introducing which-path information.
This can quantified by the well-known duality relation
\begin{equation}
    D^2 + V^2 \le 1
    \label{ineq_englert}
\end{equation}
which relates the visibility $V$ of an interferometer to the \emph{distinguishability} $D$ of the paths  Ref.~\cite{englert1996fringe}. 
When the path taken by the photon can, in principle, be perfectly known ($D = 1$), the interference visibility vanishes ($V = 0$). In the other extreme, maximal visibility ($V = 1$) requires that no which-path information can be extracted ($D = 0$). 
This is particularly relevant in the context of our experiment, as we must ensure that Bob erases any which-path information using his ancilla interferometer. 
In particular, if the paths are not overlapped properly the which-path information is not erased.
This means that a reduction of the visibility of the ancilla interferometer corresponds to a direct increase in $D$, thereby degrading $V$ and, hence, the ICO \cite{siddiqui2026complementaritydefinitecausalorder}.

We experimentally study this effect by varying the relative path length in the ancilla interferometer, reducing its visibility.
For different values of $D$\textemdash corresponding to varying path length differences and thus different interferometric visibilities\textemdash we measure the value of the causal witness. The distinguishability is obtained from the visibility $V$ of Bob's interferometer (estimated from data as in Fig.~\ref{fig:tomo_scan}.b) via $D = \sqrt{1 - V^2}$, which saturates the bound given by Eq.~\eqref{ineq_englert} in the pure-state limit. The results are shown in Fig.~\ref{fig:wit_vs_vis}.

For our minimal achievable experimental distinguishability, we obtain $\mathcal{C}_W \approx -0.305(1)$, as reported above.
As the distinguishability is increased, the system transitions from a causally non-separable regime to the classical regime of causal mixtures. 
Since equality in Eq.~\eqref{ineq_englert} is achieved only for pure states, the results presented in Fig.~\ref{fig:wit_vs_vis} are inherently lower bounded.
The residual discrepancy between the theoretical model and the experimental data is primarily due to the non-perfect purity of the experimentally reconstructed state, arising from phase fluctuations in both the quantum switch and the ancilla interferometer. 
As the distinguishability increases (i.e., as the visibility decreases), these phase fluctuations are also reduced, leading to improved agreement between the experimental results and the theoretical predictions in the causally separable regime.

\section{Conclusion}

We have introduced and experimentally demonstrated a method to perform local measurements inside a quantum switch while preserving the superposition of causal orders. 
By utilizing a time-delocalized ancilla and ensuring that any which-path information it acquires is erased, we measure the polarization of the system photon non-destructively within an indefinite causal order. 
We verified the coherence of our quantum switch via a causal witness, showing how the ICO is related to which-path information carried by the time-delocalized ancilla. 

In contrast to prior implementations of measurements in an indefinite causal order, our work allows us to treat the superposed operations as agents or parties who read their results out locally.
This is a prerequisite for many advanced ICO-based protocols, as well as a fundamental requirement for the device-independent certification of ICO. 
Thus our work takes an essential step towards a the long-standing goal of a loophole-free experimental verification indefinite causal order.
Furthermore, our scheme can be implemented within many interferometric architectures, opening new perspectives for foundational measurements outside of ICO.
Finally, our implementation explicitly highlights the time-delocalized operations as proposed by Oreshkov \cite{Oreshkov2019time}, demonstrating that although operations in photonic quantum quantum switches are delocalized they can be coherently controlled.

\section*{Data and Code availability}
All the data and code that
are necessary to replicate, verify, falsify and/or reuse
this research is available online at \cite{dataset}.

\section*{Acknowledgments}
We thank Patrick Sund, Jessica Bavaresco, Marco Tulio Quintino, Teodor Strömberg, Kyrylo Simonov, and Hector Spencer-Wood for insightful discussions.
This project has received funding from the European Union (ERC, GRAVITES, No 101071779), the European Union’s Horizon 2020 research and innovation programme under grant agreement No 899368 (EPIQUS), the European Union’s Horizon 2020 research and innovation programme under the Marie Skłodowska-Curie grant agreement No 956071 (AppQInfo) and the European Union (HORIZON Europe Research and Innovation Programme, EPIQUE, No 101135288). 
This research was funded in whole or in part by the Austrian Science Fund (FWF)[10.55776/COE1] (Quantum Science Austria), [10.55776/F71] (BeyondC) and [10.55776/FG5] (Research Group 5). 
This material is based upon work supported by the Air Force Office of Scientific Research under award number FA9550-21-1-0355 (Q-Trust) and FA8655-23-1-7063 (TIQI).
Views and opinions expressed are however those of the author(s) only and do not necessarily reflect those of the European Union or the European Research Council Executive Agency. Neither the European Union nor the granting authority can be held responsible for them.
For open access purposes, the author has applied a CC BY public copyright license to any author accepted manuscript version arising from this submission.
The financial support by the Austrian Federal Ministry of Labour and Economy, the National Foundation for Research, Technology and Development and the Christian Doppler Research Association is gratefully acknowledged.

\bibliography{bibliography}

@article{chiribella2013quantum,
  title={Quantum computations without definite causal structure},
  author={Chiribella, Giulio and D’Ariano, Giacomo Mauro and Perinotti, Paolo and Valiron, Benoit},
  journal={Physical Review A—Atomic, Molecular, and Optical Physics},
  volume={88},
  number={2},
  pages={022318},
  year={2013},
  publisher={APS}
}

@article{rozema2024experimental,
  title={Experimental aspects of indefinite causal order in quantum mechanics},
  journal={Nature Reviews Physics},
  number={8},
  pages={483--499},
  year={2024},
  publisher={Nature Publishing Group UK London}
}

@article{araujo2015witnessing,
  title={Witnessing causal nonseparability},
  author={Ara{\'u}jo, Mateus and Branciard, Cyril and Costa, Fabio and Feix, Adrien and Giarmatzi, Christina and Brukner, {\v{C}}aslav},
  journal={New Journal of Physics},
  volume={17},
  number={10},
  pages={102001},
  year={2015},
  publisher={IOP Publishing}
}

@article{lorenzo2015,
       author = {{Procopio}, Lorenzo M. and {Moqanaki}, Amir and {Ara{\'u}jo}, Mateus and {Costa}, Fabio and {Alonso Calafell}, Irati and {Dowd}, Emma G. and {Hamel}, Deny R. and {Rozema}, Lee A. and {Brukner}, {\v{C}}aslav and {Walther}, Philip},
        title = "{Experimental superposition of orders of quantum gates}",
      journal = {Nature Communications},
         year = 2015,
        month = aug,
       volume = {6},
          eid = {7913},
        pages = {7913},
          doi = {10.1038/ncomms8913},
archivePrefix = {arXiv},
       eprint = {1412.4006},
 primaryClass = {quant-ph},
       adsurl = {https://ui.adsabs.harvard.edu/abs/2015NatCo...6.7913P}
}

@article{bavaresco19,
  doi = {10.22331/q-2019-08-19-176},
  url = {https://doi.org/10.22331/q-2019-08-19-176},
  title = {Semi-device-independent certification of indefinite causal order},
  author = {Bavaresco, Jessica and Ara{\'{u}}jo, Mateus and Brukner, {\v{C}}aslav and Quintino, Marco T{\'{u}}lio},
  journal = {{Quantum}},
  issn = {2521-327X},
  publisher = {{Verein zur F{\"{o}}rderung des Open Access Publizierens in den Quantenwissenschaften}},
  volume = {3},
  pages = {176},
  year = {2019},
  archivePrefix = {arXiv},
  eprint = {1903.10526},
  primaryClass = {quant-ph}  
}

@article{englert1996fringe,
  title={Fringe visibility and which-way information: An inequality},
  author={Englert, Berthold-Georg},
  journal={Physical review letters},
  volume={77},
  number={11},
  pages={2154},
  year={1996},
  publisher={APS}
}

@article{spencer2025indefinite,
  title={Indefinite causal key distribution},
  author={Spencer-Wood, Hector},
  journal={Journal of Physics A: Mathematical and Theoretical},
  year={2025}
}

@ARTICLE{Oreshkov2019time,
       author = {{Oreshkov}, Ognyan},
        title = "{Time-delocalized quantum subsystems and operations: on the existence of processes with indefinite causal structure in quantum mechanics}",
      journal = {Quantum},
         year = 2019,
        month = dec,
       volume = {3},
        pages = {206},
          doi = {10.22331/q-2019-12-02-206},
archivePrefix = {arXiv},
       eprint = {1801.07594},
 primaryClass = {quant-ph},
       adsurl = {https://ui.adsabs.harvard.edu/abs/2019Quant...3..206O}
}

@article{oreshkov2012quantum,
       author = {{Oreshkov}, Ognyan and {Costa}, Fabio and {Brukner}, {\v{C}}aslav},
        title = "{Quantum correlations with no causal order}",
      journal = {Nature Communications},
         year = 2012,
        month = oct,
       volume = {3},
          eid = {1092},
        pages = {1092},
          doi = {10.1038/ncomms2076},
archivePrefix = {arXiv},
       eprint = {1105.4464},
 primaryClass = {quant-ph},
       adsurl = {https://ui.adsabs.harvard.edu/abs/2012NatCo...3.1092O}
}

@article{baumeler2014maximal,
  title={Maximal incompatibility of locally classical behavior and global causal order in multiparty scenarios},
  author={Baumeler, {\"A}min and Feix, Adrien and Wolf, Stefan},
  journal={Physical Review A},
  volume={90},
  number={4},
  pages={042106},
  year={2014},
  publisher={APS}
}

@article{baumeler2016space,
  title={The space of logically consistent classical processes without causal order},
  author={Baumeler, {\"A}min and Wolf, Stefan},
  journal={New Journal of Physics},
  volume={18},
  number={1},
  pages={013036},
  year={2016},
  publisher={IOP Publishing}
}

@article{wechs2021quantum,
  title={Quantum circuits with classical versus quantum control of causal order},
  author={Wechs, Julian and Dourdent, Hippolyte and Abbott, Alastair A and Branciard, Cyril},
  journal={PRX Quantum},
  volume={2},
  number={3},
  pages={030335},
  year={2021},
  publisher={APS}
}

@ARTICLE{bavaresco2022UnitaryChannelDiscrimination,
       author = {{Bavaresco}, Jessica and {Murao}, Mio and {Quintino}, Marco T{\'u}lio},
        title = "{Unitary channel discrimination beyond group structures: Advantages of sequential and indefinite-causal-order strategies}",
      journal = {Journal of Mathematical Physics},
         year = 2022,
        month = apr,
       volume = {63},
       number = {4},
          eid = {042203},
        pages = {042203},
          doi = {10.1063/5.0075919},
archivePrefix = {arXiv},
       eprint = {2105.13369},
 primaryClass = {quant-ph},
       adsurl = {https://ui.adsabs.harvard.edu/abs/2022JMP....63d2203B}
}

@Article{Araujo2014,
  author       = {Mateus Ara{\'{u}}jo and Adrien Feix and Fabio Costa and {\v{C}}aslav Brukner},
  date         = {2014-09},
  journal = {New J.~Phys.},
  title        = {Quantum circuits cannot control unknown operations},
  doi          = {10.1088/1367-2630/16/9/093026},
  number       = {9},
  pages        = {093026},
  volume       = {16},
  publisher    = {{IOP} Publishing},
}

@article{taddei2021computational,
  title = {Computational Advantage from the Quantum Superposition of Multiple Temporal Orders of Photonic Gates},
  author = {Taddei, M\'arcio M. and Cari\~ne, Jaime and Mart\'{\i}nez, Daniel and Garc\'{\i}a, Tania and Guerrero, Nayda and Abbott, Alastair A. and Ara\'ujo, Mateus and Branciard, Cyril and G\'omez, Esteban S. and Walborn, Stephen P. and Aolita, Leandro and Lima, Gustavo},
  journal = {PRX Quantum},
  volume = {2},
  issue = {1},
  pages = {010320},
  numpages = {16},
  year = {2021},
  month = {Feb},
  publisher = {American Physical Society},
  doi = {10.1103/PRXQuantum.2.010320},
archivePrefix = {arXiv},
       eprint = {2002.07817},
 primaryClass = {quant-ph},
       adsurl = {https://ui.adsabs.harvard.edu/abs/2020arXiv200207817T}
}

@ARTICLE{renner2022advantage,
       author = {{Renner}, Martin J. and {Brukner}, {\v{C}}aslav},
        title = "{Computational Advantage from a Quantum Superposition of Qubit Gate Orders}",
      journal = {Phys. Rev. Lett.},
         year = 2022,
        month = jun,
       volume = {128},
       number = {23},
          eid = {230503},
        pages = {230503},
          doi = {10.1103/PhysRevLett.128.230503},
archivePrefix = {arXiv},
       eprint = {2112.14541},
 primaryClass = {quant-ph},
       adsurl = {https://ui.adsabs.harvard.edu/abs/2022PhRvL.128w0503R}
}

@INPROCEEDINGS{baumeler13,
author={Baumeler, {\"A} . and Wolf, S.},
booktitle={Information Theory (ISIT), 2014 IEEE International Symposium on},
title={Perfect signaling among three parties violating predefined causal order},
year={2014},
month={June},
pages={526-530},
doi={10.1109/ISIT.2014.6874888},
archivePrefix = "arXiv",
   eprint = {1312.5916},
 primaryClass = "quant-ph"
}

@Article{feixquantum2015,
  author    = {Feix, Adrien and Ara\'ujo, Mateus and Brukner, {\v C}aslav},
  title     = {Quantum superposition of the order of parties as a communication resource},
  journal   = {Phys. Rev. A},
  year      = {2015},
  volume    = {92},
  issue     = {5},
  month     = {Nov},
  pages     = {052326},
  doi       = {10.1103/PhysRevA.92.052326},
  numpages  = {5},
  publisher = {American Physical Society},
}

@Article{Guerin2016,
  author       = {Gu\'erin, Philippe Allard and Feix, Adrien and Ara\'ujo, Mateus and Brukner, {\v C}aslav},
  date         = {2016-09},
  journal = {Phys. Rev. Lett.},
  title        = {Exponential Communication Complexity Advantage from Quantum Superposition of the Direction of Communication},
  doi          = {10.1103/PhysRevLett.117.100502},
  issue        = {10},
  pages        = {100502},
  volume       = {117},
  numpages     = {5},
  publisher    = {American Physical Society},
}

@ARTICLE{Felce2021IBMswitch,
       author = {{Felce}, David and {Vedral}, Vlatko and {Tennie}, Felix},
        title = "{Refrigeration with Indefinite Causal Orders on a Cloud Quantum Computer}",
      journal = {arXiv e-prints},
         year = 2021,
        month = jul,
          eid = {arXiv:2107.12413},
        pages = {arXiv:2107.12413},
          doi = {10.48550/arXiv.2107.12413},
archivePrefix = {arXiv},
       eprint = {2107.12413},
 primaryClass = {quant-ph},
       adsurl = {https://ui.adsabs.harvard.edu/abs/2021arXiv210712413F}
}

@Article{Guha2020,
  author      = {Tamal Guha and Mir Alimuddin and Preeti Parashar},
  date        = {2020-03-03},
  title       = {Thermodynamic advancement in the causally inseparable occurrence of thermal maps},
  eprint      = {arXiv:2003.01464},
  eprintclass = {quant-ph},
  journal  = {arXiv},
}

@ARTICLE{Simonov2022WorkExtraction,
       author = {{Simonov}, Kyrylo and {Francica}, Gianluca and {Guarnieri}, Giacomo and {Paternostro}, Mauro},
        title = "{Work extraction from coherently activated maps via quantum switch}",
      journal = {Phys. Rev. A},
         year = 2022,
        month = mar,
       volume = {105},
       number = {3},
          eid = {032217},
        pages = {032217},
          doi = {10.1103/PhysRevA.105.032217},
       adsurl = {https://ui.adsabs.harvard.edu/abs/2022PhRvA.105c2217S}
}

@ARTICLE{Frey2019depolarizingChannelIdentification,
       author = {{Frey}, Michael},
        title = "{Indefinite causal order aids quantum depolarizing channel identification}",
      journal = {Quantum Information Processing},
         year = 2019,
        month = apr,
       volume = {18},
       number = {4},
          eid = {96},
        pages = {96},
          doi = {10.1007/s11128-019-2186-9},
       adsurl = {https://ui.adsabs.harvard.edu/abs/2019QuIP...18...96F}
}

@Article{ChapeauBlondeau2022,
  author       = {François Chapeau-Blondeau},
  date         = {2022},
  journal = {Physics Letters A},
  title        = {Indefinite causal order for quantum metrology with quantum thermal noise},
  doi          = {https://doi.org/10.1016/j.physleta.2022.128300},
  issn         = {0375-9601},
  pages        = {128300},
  url          = {https://www.sciencedirect.com/science/article/pii/S0375960122003826},
  volume       = {447},
}

@article{rubino2017ExperimentalVerification,
       author = {{Rubino}, Giulia and {Rozema}, Lee A. and {Feix}, Adrien and {Ara{\'u}jo}, Mateus and {Zeuner}, Jonas M. and {Procopio}, Lorenzo M. and {Brukner}, {\v{C}}aslav and {Walther}, Philip},
        title = "{Experimental verification of an indefinite causal order}",
      journal = {Science Advances},
         year = 2017,
        month = mar,
       volume = {3},
       number = {3},
        pages = {e1602589},
          doi = {10.1126/sciadv.1602589},
archivePrefix = {arXiv},
       eprint = {1608.01683},
 primaryClass = {quant-ph},
       adsurl = {https://ui.adsabs.harvard.edu/abs/2017SciA....3E2589R}
}

@article{Rubino2022experimentalEntanglement,
  doi = {10.22331/q-2022-01-11-621},
  title = {Experimental entanglement of temporal order},
  author = {Rubino, Giulia and Rozema, Lee A. and Massa, Francesco and Ara{\'{u}}jo, Mateus and Zych, Magdalena and Brukner, {\v{C}}aslav and Walther, Philip},
  journal = {{Quantum}},
  issn = {2521-327X},
  publisher = {{Verein zur F{\"{o}}rderung des Open Access Publizierens in den Quantenwissenschaften}},
  volume = {6},
  pages = {621},
  archivePrefix = {arXiv},
       eprint = {1712.06884},
 primaryClass = {quant-ph},
  month = jan,
  year = {2022}
}

@article{Rubino2021Communication,
       author = {{Rubino}, Giulia and {Rozema}, Lee A. and {Ebler}, Daniel and {Kristj{\'a}nsson}, Hl{\'e}r and {Salek}, Sina and {Allard Gu{\'e}rin}, Philippe and {Abbott}, Alastair A. and {Branciard}, Cyril and {Brukner}, {\v{C}}aslav and {Chiribella}, Giulio and {Walther}, Philip},
        title = "{Experimental quantum communication enhancement by superposing trajectories}",
      journal = {Physical Review Research},
         year = 2021,
        month = jan,
       volume = {3},
       number = {1},
          eid = {013093},
        pages = {013093},
          doi = {10.1103/PhysRevResearch.3.013093},
archivePrefix = {arXiv},
       eprint = {2007.05005},
 primaryClass = {quant-ph},
       adsurl = {https://ui.adsabs.harvard.edu/abs/2021PhRvR...3a3093R}
}

@article{guo2020experimental,
title = {Experimental Transmission of Quantum Information Using a Superposition of Causal Orders},
  author = {Guo, Yu and Hu, Xiao-Min and Hou, Zhi-Bo and Cao, Huan and Cui, Jin-Ming and Liu, Bi-Heng and Huang, Yun-Feng and Li, Chuan-Feng and Guo, Guang-Can and Chiribella, Giulio},
  journal = {Phys. Rev. Lett.},
  volume = {124},
  issue = {3},
  pages = {030502},
  numpages = {6},
    publisher = {American Physical Society},
  doi = {10.1103/PhysRevLett.124.030502},
  year = {2020},
  month = {Jan},
archivePrefix = {arXiv},
       eprint = {1811.07526},
 primaryClass = {quant-ph},
       adsurl = {https://ui.adsabs.harvard.edu/abs/2018arXiv181107526G}
}

@article{cao2022Semideviceindependent,
       author = {{Cao}, Huan and {Bavaresco}, Jessica and {Wang}, Ning-Ning and {Rozema}, Lee A. and {Zhang}, Chao and {Huang}, Yun-Feng and {Liu}, Bi-Heng and {Li}, Chuan-Feng and {Guo}, Guang-Can and {Walther}, Philip},
        title = "{Semi-device-independent certification of indefinite causal order in a photonic quantum switch}",
      journal = {Optica},
         year = 2023,
        month = may,
       volume = {10},
       number = {5},
        pages = {561},
          doi = {10.1364/OPTICA.483876},
archivePrefix = {arXiv},
       eprint = {2202.05346},
 primaryClass = {quant-ph},
       adsurl = {https://ui.adsabs.harvard.edu/abs/2023Optic..10..561C}
}

@article{goswami2018Indefinite,
  title={Indefinite causal order in a quantum switch},
  author={Goswami, K and Giarmatzi, Christina and Kewming, M and Costa, Fabio and Branciard, Cyril and Romero, Jacquiline and White, A. G.},
  journal={Phys. Rev. Lett.},
  volume={121},
  number={9},
  pages={090503},
  year={2018},
  publisher={APS},
  DOI={10.1103/PhysRevLett.121.090503}
}

@article{goswami2020IncreasingCommunication,
  title = {Increasing communication capacity via superposition of order},
  author = {Goswami, K. and Cao, Y. and Paz-Silva, G. A. and Romero, J. and White, A. G.},
  journal = {Phys. Rev. Research},
  volume = {2},
  issue = {3},
  pages = {033292},
  numpages = {9},
  year = {2020},
  month = {Aug},
  publisher = {American Physical Society},
  doi = {10.1103/PhysRevResearch.2.033292}
}

@article{wei2019experimentalCommunication,
       author = {{Wei}, Kejin and {Tischler}, Nora and {Zhao}, Si-Ran and {Li}, Yu-Huai and {Arrazola}, Juan Miguel and {Liu}, Yang and {Zhang}, Weijun and {Li}, Hao and {You}, Lixing and {Wang}, Zhen and {Chen}, Yu-Ao and {Sanders}, Barry C. and {Zhang}, Qiang and {Pryde}, Geoff J. and {Xu}, Feihu and {Pan}, Jian-Wei},
        title = "{Experimental Quantum Switching for Exponentially Superior Quantum Communication Complexity}",
      journal = {Phys. Rev. Lett.},
         year = 2019,
        month = mar,
       volume = {122},
       number = {12},
          eid = {120504},
        pages = {120504},
          doi = {10.1103/PhysRevLett.122.120504},
archivePrefix = {arXiv},
       eprint = {1810.10238},
 primaryClass = {quant-ph},
       adsurl = {https://ui.adsabs.harvard.edu/abs/2019PhRvL.122l0504W}
}

@article{Antesberger2023tomography,
  title = {Higher-Order Process Matrix Tomography of a Passively-Stable Quantum Switch},
  author = {Antesberger, Michael and Quintino, Marco T\'ulio and Walther, Philip and Rozema, Lee A.},
  journal = {PRX Quantum},
  volume = {5},
  issue = {1},
  pages = {010325},
  numpages = {22},
  year = {2024},
  month = {Feb},
  publisher = {American Physical Society},
  doi = {10.1103/PRXQuantum.5.010325},
  url = {https://link.aps.org/doi/10.1103/PRXQuantum.5.010325}
}

@ARTICLE{cao2022quantumSimulation,
       author = {{Cao}, Huan and {Wang}, Ning-Ning and {Jia}, Zhian and {Zhang}, Chao and {Guo}, Yu and {Liu}, Bi-Heng and {Huang}, Yun-Feng and {Li}, Chuan-Feng and {Guo}, Guang-Can},
        title = "{Quantum simulation of indefinite causal order induced quantum refrigeration}",
      journal = {Physical Review Research},
         year = 2022,
        month = aug,
       volume = {4},
       number = {3},
          eid = {L032029},
        pages = {L032029},
          doi = {10.1103/PhysRevResearch.4.L032029},
archivePrefix = {arXiv},
       eprint = {2101.07979},
 primaryClass = {quant-ph},
       adsurl = {https://ui.adsabs.harvard.edu/abs/2022PhRvR...4c2029C}
}

@article{zhu2023prl,
  title = {Charging Quantum Batteries via Indefinite Causal Order: Theory and Experiment},
  author = {Zhu, Gaoyan and Chen, Yuanbo and Hasegawa, Yoshihiko and Xue, Peng},
  journal = {Phys. Rev. Lett.},
  volume = {131},
  issue = {24},
  pages = {240401},
  numpages = {7},
  year = {2023},
  month = {Dec},
  publisher = {American Physical Society},
  doi = {10.1103/PhysRevLett.131.240401},
  url = {https://link.aps.org/doi/10.1103/PhysRevLett.131.240401}
}

@article{Min23,
  title = {Noisy quantum parameter estimation with indefinite causal order},
  author = {An, Min and Ru, Shihao and Wang, Yunlong and Yang, Yu and Wang, Feiran and Zhang, Pei and Li, Fuli},
  journal = {Phys. Rev. A},
  volume = {109},
  issue = {1},
  pages = {012603},
  numpages = {19},
  year = {2024},
  month = {Jan},
  publisher = {American Physical Society},
  doi = {10.1103/PhysRevA.109.012603},
  url = {https://link.aps.org/doi/10.1103/PhysRevA.109.012603}
}

@article{yin2023experimental,
  title={Experimental super-Heisenberg quantum metrology with indefinite gate order},
  author={Yin, Peng and Zhao, Xiaobin and Yang, Yuxiang and Guo, Yu and Zhang, Wen-Hao and Li, Gong-Chu and Han, Yong-Jian and Liu, Bi-Heng and Xu, Jin-Shi and Chiribella, Giulio and others},
  journal={Nature Physics},
  volume={19},
  number={8},
  pages={1122--1127},
  year={2023},
  publisher={Nature Publishing Group UK London}
}

@article{Ning2023MeasuringIncompatibility,
  title = {Measuring Incompatibility and Clustering Quantum Observables with a Quantum Switch},
  author = {Gao, Ning and Li, Dantong and Mishra, Anchit and Yan, Junchen and Simonov, Kyrylo and Chiribella, Giulio},
  journal = {Phys. Rev. Lett.},
  volume = {130},
  issue = {17},
  pages = {170201},
  numpages = {6},
  year = {2023},
  month = {Apr},
  publisher = {American Physical Society},
  doi = {10.1103/PhysRevLett.130.170201},
  url = {https://link.aps.org/doi/10.1103/PhysRevLett.130.170201}
}

@article{van2023device,
  title={Device-independent certification of indefinite causal order in the quantum switch},
  author={Van Der Lugt, Tein and Barrett, Jonathan and Chiribella, Giulio},
  journal={Nature Communications},
  volume={14},
  number={1},
  pages={5811},
  year={2023},
  publisher={Nature Publishing Group UK London}
}

@article{richter2025towards,
  title={Towards an Experimental Device-Independent Verification of Indefinite Causal Order},
  author={Richter, Carla and Antesberger, Michael and Cao, Huan and Walther, Philip and Rozema, Lee A},
  journal={arXiv preprint arXiv:2506.16949},
  year={2025}
}

@Article{Zhao2020,
  author       = {Zhao, Xiaobin and Yang, Yuxiang and Chiribella, Giulio},
  date         = {2020-05},
  journal = {Phys. Rev. Lett.},
  title        = {Quantum Metrology with Indefinite Causal Order},
  doi          = {10.1103/PhysRevLett.124.190503},
  issue        = {19},
  pages        = {190503},
  url          = {https://link.aps.org/doi/10.1103/PhysRevLett.124.190503},
  volume       = {124},
  numpages     = {6},
  publisher    = {American Physical Society},
}

@article{pryde2005measurement,
  title={Measurement of quantum weak values of photon polarization},
  author={Pryde, GJ and O’Brien, JL and White, AG and Ralph, TC and Wiseman, HM},
  journal={Physical review letters},
  volume={94},
  number={22},
  pages={220405},
  year={2005},
  publisher={APS}
}

@article{ralph2002linear,
  title={Linear optical controlled-NOT gate in the coincidence basis},
  author={Ralph, Timothy C and Langford, Nathan K and Bell, TB and White, AG},
  journal={Physical Review A},
  volume={65},
  number={6},
  pages={062324},
  year={2002},
  publisher={APS}
}

@article{pittman2001probabilistic,
  title={Probabilistic quantum logic operations using polarizing beam splitters},
  author={Pittman, TB and Jacobs, BC and Franson, JD},
  journal={Physical Review A},
  volume={64},
  number={6},
  pages={062311},
  year={2001},
  publisher={APS}
}

@ARTICLE{KLM,
       author = {{Knill}, E. and {Laflamme}, R. and {Milburn}, G.~J.},
        title = "{A scheme for efficient quantum computation with linear optics}",
      journal = {\nat},
         year = 2001,
        month = jan,
       volume = {409},
       number = {6816},
        pages = {46-52},
          doi = {10.1038/35051009},
       adsurl = {https://ui.adsabs.harvard.edu/abs/2001Natur.409...46K}
}

@ARTICLE{pryde2004measuring,
       author = {{Pryde}, G.~J. and {O'Brien}, J.~L. and {White}, A.~G. and {Bartlett}, S.~D. and {Ralph}, T.~C.},
        title = "{Measuring a Photonic Qubit without Destroying It}",
      journal = {\prl},
         year = 2004,
        month = may,
       volume = {92},
       number = {19},
          eid = {190402},
        pages = {190402},
          doi = {10.1103/PhysRevLett.92.190402},
archivePrefix = {arXiv},
       eprint = {quant-ph/0312048},
 primaryClass = {quant-ph},
       adsurl = {https://ui.adsabs.harvard.edu/abs/2004PhRvL..92s0402P}
}

@ARTICLE{Purves2021CannotViolate,
       author = {{Purves}, Tom and {Short}, Anthony J.},
        title = "{Quantum Theory Cannot Violate a Causal Inequality}",
      journal = {Phys. Rev. Lett.},
         year = 2021,
        month = sep,
       volume = {127},
       number = {11},
          eid = {110402},
        pages = {110402},
          doi = {10.1103/PhysRevLett.127.110402},
archivePrefix = {arXiv},
       eprint = {2101.09107},
 primaryClass = {quant-ph},
       adsurl = {https://ui.adsabs.harvard.edu/abs/2021PhRvL.127k0402P}
}

@article{barz2014two,
  title={A two-qubit photonic quantum processor and its application to solving systems of linear equations},
  author={Barz, Stefanie and Kassal, Ivan and Ringbauer, Martin and Lipp, Yannick Ole and Daki{\'c}, Borivoje and Aspuru-Guzik, Al{\'a}n and Walther, Philip},
  journal={Scientific reports},
  volume={4},
  number={1},
  pages={6115},
  year={2014},
  publisher={Nature Publishing Group UK London}
}

@article{lesniak2026bipartite,
  title={A Bipartite Quantum Key Distribution Protocol Based on Indefinite Causal Order},
  author={Le{\'s}niak, Mateusz and Kukulski, Ryszard and Lewandowska, Paulina and Rajchel-Mieldzio{\'c}, Grzegorz and Wro{\'n}ski, Micha{\l}},
  journal={arXiv preprint arXiv:2603.08204},
  year={2026}
}

@misc{siddiqui2026complementaritydefinitecausalorder,
      title={Complementarity Beyond Definite Causal Order}, 
      author={Mohd Asad Siddiqui and Md Qutubuddin and Tabish Qureshi},
      year={2026},
      eprint={2603.27780},
      archivePrefix={arXiv},
      primaryClass={quant-ph},
      url={https://arxiv.org/abs/2603.27780}, 
}

@misc{dataset,
    title = {Time-Delocalized Local Measurements in an Indefinite Causal Order, 10.5281/zenodo.19334935 (2026)},
    author = {Valibouse, Yann and Cladera-Rossell{\'o}, Mart{\'i} and Antesberger, Michael and Walther, Philip and Rozema, Lee A},
    doi = {10.5281/zenodo.19334935}
}

\newpage
\appendix

\section{Measurement inside the Quantum Switch}
To perform a measurement of the system polarization, we use an ancilla photon inspired by the von Neumann measurement model. In our experiment, we fix the state of the ancilla photon to
\begin{equation}
    \ket{D}_a = \tfrac{1}{\sqrt{2}}(\ket{H}_a + \ket{V}_a)
\end{equation}
while the system photon can be in an arbitrary pure state
\begin{equation}
    \ket{\psi}_s = \alpha \ket{H}_s + \beta \ket{V}_s 
\end{equation}
After the PBS and post-selection on events where there is one photon in each output port, we obtain the state
\begin{equation}
    \ket{\psi'} = \alpha \ket{H}_s \ket{H}_a + \beta \ket{V}_s \ket{V}_a 
\end{equation}
which corresponds to the measurement interaction discussed above.

We now consider implementing this measurement inside the quantum switch. Starting from an entangled pair of photons generated by an SPDC source,
$(\ket{H}_s\ket{H}_a+\ket{V}_s\ket{V}_a)/\sqrt{2}$, and sending each photon to a different PBS, we obtain the hyperentangled state
\begin{equation}
    (\ket{H,0}_s\ket{H,0}_a+\ket{V,1}_s\ket{V,1}_a)/\sqrt{2}
\end{equation}
where $0,1$ denote the path of the system photon (i.e.\ the paths corresponding to the two causal orders in the quantum switch) and the associated paths of Bob's ancilla photon.

We then set the same polarization on both paths of the system photon (i.e.\ for both causal orders) using a quarter-wave plate and a half-wave plate, allowing the preparation of any pure polarization state. Furthermore, we initialize the polarization of the ancilla photon on both paths using a half-wave plate.\\

The initial state at the beginning of the quantum switch is therefore
\begin{align}
    \ket{\psi}_{\mathrm{in}} = &\frac{1}{2}\bigg[(\alpha\ket{H,0}_{s,c} + \beta \ket{V,0}_{s,c})\otimes \ket{D,0}_{a,p} \notag \\
    &+ (\alpha\ket{H,1}_{s,c} + \beta \ket{V,1}_{s,c}) \otimes \ket{D,1}_{a,p} \bigg]
\end{align}
with $|\alpha|^2+|\beta|^2=1$.\\

After the system evolves inside the quantum switch—for instance, passing through Alice's unitary first in one path—the polarization arriving at Bob's station may differ between the two causal orders,
\begin{align}
    \ket{\psi}_{\mathrm{Bob_{in}}} = &\frac{1}{2}\bigg[(\alpha_{A\to B}\ket{H,0}_{s,c} + \beta_{A\to B} \ket{V,0}_{s,c})\otimes \ket{D,0}_{a,p} \notag\\
    &+ (\alpha_{B\to A}\ket{H,1}_{s,c} + \beta_{B\to A} \ket{V,1}_{s,c}) \otimes \ket{D,1}_{a,p} \bigg]
\end{align}
with $|\alpha_i|^2+|\beta_i|^2=1$ for $i=A\to B, B\to A$.

After Bob's PBSs and post-selection on events where one photon is detected in the ancilla paths and one photon remains inside the quantum switch, we obtain the state
\begin{align}
    \ket{\psi}_{\mathrm{Bob}} = &\frac{1}{2}\bigg[\left(\alpha_{A\to B}\ket{HH}_{s,a} + \beta_{A\to B} \ket{VV}_{s,a}\right)\ket{0,0}_{c,p} \notag\\
   & + \left(\alpha_{B\to A}\ket{HH}_{s,a} + \beta_{B\to A} \ket{VV}_{s,a}\right)\ket{1,1}_{c,p}\bigg] 
   \label{bob_state}
\end{align}

This entangled state already allows the measurement of the system polarization by measuring the ancilla photon. However, in this state the paths are also correlated, meaning that measuring the ancilla would collapse the path coherence of the system and therefore destroy the indefinite causal order. 

To avoid this, we recombine the ancilla paths on a beamsplitter after Bob's measurement interaction, which leads to the state
\begin{align}
    \ket{\psi}_{\mathrm{Bob}} = \frac{1}{2\sqrt{2}}&\bigg[\alpha_{A\to B}\ket{HH}_{s,a}\ket{0}_{c} (\ket{0}_p + \ket{1}_p)\notag \\
    &+ \beta_{A\to B} \ket{VV}_{s,a}\ket{0}_{c} (\ket{0}_p + \ket{1}_p) \notag\\
   & + \alpha_{B\to A}\ket{HH}_{s,a}\ket{1}_{c} (\ket{0}_p - \ket{1}_p) \notag\\
   &+ \beta_{B\to A} \ket{VV}_{s,a}\ket{1}_{c} (\ket{0}_p - \ket{1}_p)\bigg] 
   \label{bob_state}
\end{align}

This procedure allows the polarization to be measured without collapsing the path coherence. After the final beamsplitter, due to the sign difference between the output ports of the ancilla beamsplitter, the output ports of the quantum switch become correlated with the output ports of the ancilla. However, because the ancilla path information has been erased, this correlation does not affect the path coherence inside the quantum switch.

\section{Process Matrix framework}
To describe indefinite causal structures, we rely on the \emph{process matrix framework}. A process matrix plays a role analogous to that of a density matrix: while a density matrix encodes the information of a quantum \emph{state}, a process matrix characterizes an ensemble of \emph{quantum processes} connecting local laboratories.\\
We consider two parties, Alice and Bob, who perform local quantum operations in their respective laboratories. Alice’s unitary operation is denoted by $U^A$, and Bob’s measurement operation by $M^B$. The process matrix $W$ specifies how quantum systems are transferred between these laboratories and is therefore independent of the particular local operations chosen by Alice and Bob.\\
We denote Alice’s and Bob’s respective choices of operations by $x$ and $y$ and Bob's operation outcome $b$. The preparation of the input state is labeled by $z$, while the outcome of a final detection operation is denoted by $d$. With these definitions, the joint probability of obtaining outcomes $b$, and $d$, given the settings $x$, $y$, and $z$, is given by
\begin{equation}
p(b,d \mid x,y,z)
= \mathrm{Tr}\!\left[
\rho^{(\mathrm{in})}_z
\otimes U^{A}_{x}
\otimes M^{B}_{b|y}
\otimes D^{(\mathrm{out})}_d
\, W
\right]
\end{equation}
This probability distribution satisfies the normalization condition
\begin{equation}
\sum_{b,d} p(b,d \mid x,y,z) = 1
\quad \forall\, x,y,z,
\end{equation}
where $W$ is the process matrix.\\
In the case of a definite causal order, one can define two classes of process matrices: $W^{A\to B}$, describing processes in which Alice’s operation precedes Bob’s, and $W^{B\to A}$, describing processes in which Bob acts before Alice. A process is said to be \emph{causally separable} if it can be written as a mixture of these two definite-order processes,
\begin{equation}
W^{\mathrm{sep}} := p\, W^{A\to B} + (1-p)\, W^{B\to A}
\label{w_sep}
\end{equation}
with $0 \leq p \leq 1$. Analogously to a classical mixture of quantum states, the process $W^{\mathrm{sep}}$ corresponds to a situation in which the order $A \to B$ occurs with probability $p$, and the order $B \to A$ occurs with probability $1-p$.\\
Processes that cannot be written in this form are called \emph{causally non-separable}. To distinguish between causally separable and causally non-separable processes, we use the concept of \emph{causal witness} ($\mathcal{C}_W$) analogous to an entanglement witness. As shown in Ref.~\cite{araujo2015witnessing}, for any causally non-separable process matrix $W^{\mathrm{non\text{-}sep}}$, there exists a Hermitian operator $S$, called a causal witness matrix, such that
\begin{equation}
\mathcal{C}_W:=\mathrm{Tr}\!\left( S W^{\mathrm{non\text{-}sep}} \right) < 0
\end{equation}
while
\begin{equation}
\mathcal{C}_W:=\mathrm{Tr}\!\left( S W^{\mathrm{sep}} \right) \geq 0
\end{equation}
for all causally separable process matrices $W^{\mathrm{sep}}$. Operationally, a causal witness probes correlations that cannot be reproduced by any probabilistic mixture of definite causal orders. If the observed statistics can be explained by a classical random choice between the orders $A \to B$ and $B \to A$, then the expectation value of any causal witness is positive. A negative value of $\mathcal{C}_W$ therefore certifies that no underlying definite causal order can account for the observed correlations. The procedure to compute an optimal causal witness is detailed in Ref.~\cite{araujo2015witnessing}. Importantly, the expectation value of a causal witness can be expressed directly in terms of experimentally accessible probabilities as
\begin{equation}
\mathcal{C}_W
= \sum_{\substack{b,d \\ x,y,z}}
\alpha_{b,d,x,y,z}\,
p(b,d \mid x,y,z)
\label{causal_witness}
\end{equation}
where the coefficients $\alpha_{b,d,x,y,z}$ depend on the specific witness $S$.
\\
In the specific case of the \emph{quantum switch}, the process matrix routes the input state to Alice and Bob in a coherent superposition of causal orders. The output of Alice is coherently connected to the input of Bob and vice versa, before the two possible evolutions are recombined. In our analysis, we use the process matrix $W_{\mathrm{SWITCH}}$ reported in
Ref.~\cite{rubino2017ExperimentalVerification}.\\

\begin{figure}
    \centering
    \includegraphics[width=\textwidth]{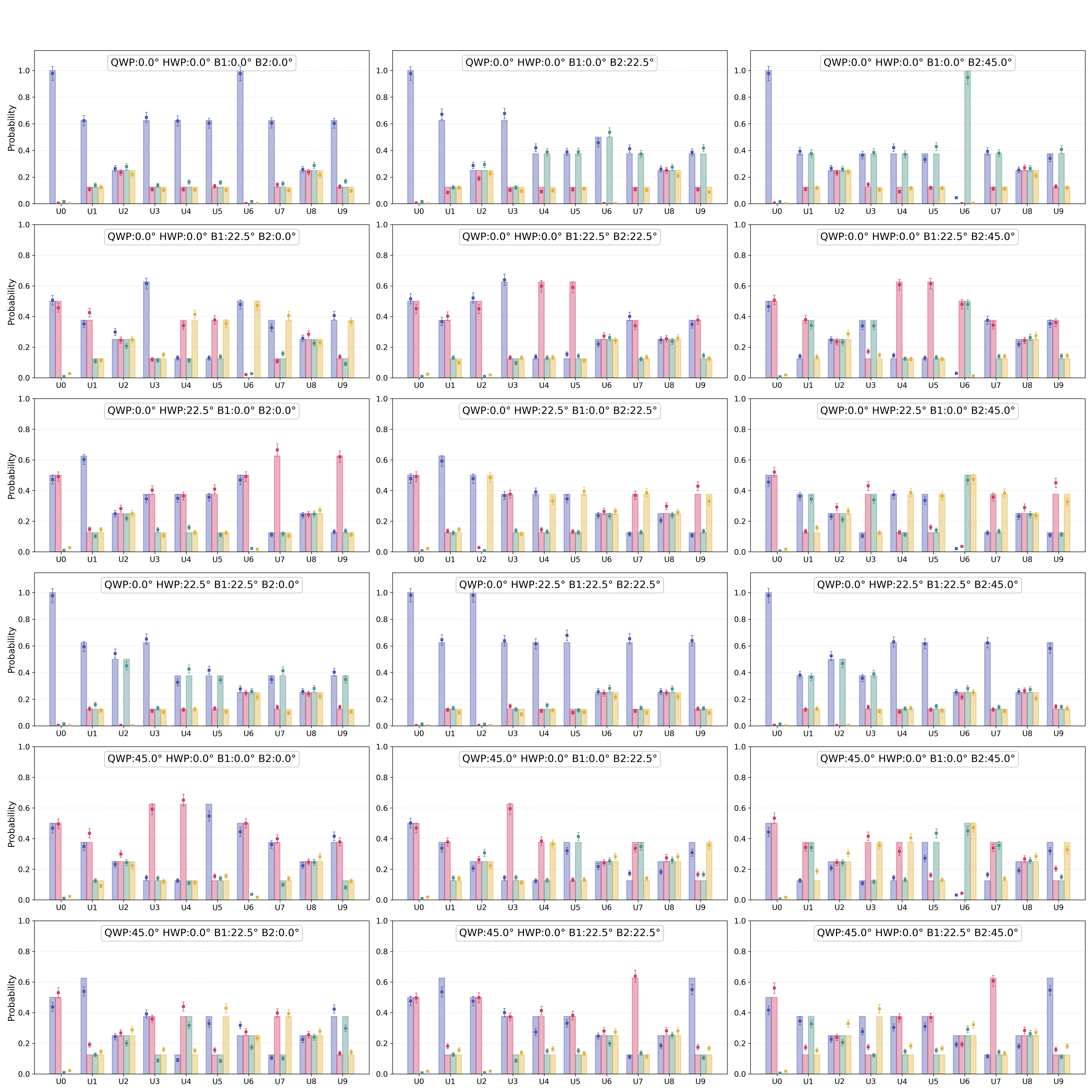}
    \caption{Probabilities for each possible outputs for Alice's ten unitaries operation for all $z, y$ settings.}
    \label{fig:barplot}
\end{figure}

For our experiment, we can prepare the input state as $\ket{H}, \ket{D}, \ket{L}$ ($z$ setting). We let Bob measure and reprepare in the $Z-$ and $X-$ basis ($y$ setting). Alice performs a set of 10 unitaries ($x$ setting). The probibily measured in the joined probability of Bob measuring the polarization result $b$ with the control qubit probability $d$. This leads to possible 180 differents experimental settings described in table~\ref{tab:waveplate_angles} whose results are shown on Fig.~\ref{fig:barplot}

\begin{table*}[h]
\centering
\caption{Set of wave plate angles. A list of all wave plate angles used to perform the operators. All combinations of these settings were used, resulting in 180 measurement settings.}
\label{tab:waveplate_angles}
\begin{tabular}{llll}
\hline\hline
\textbf{Input state} & \textbf{Bob's measurement} & \textbf{Bob's repreparation} & \textbf{Alice's unitary} \\
\textbf{(QWP, HWP)} & \textbf{(HWP)} & \textbf{(HWP)} & \textbf{(QWP, HWP, QWP)} \\
\hline\hline
(1) $0^\circ_\text{QWP}$, $0^\circ_\text{HWP}$
  & (1) $0^\circ_\text{HWP}$
  & (1) $0^\circ_\text{HWP}$
  & (1) $0^\circ_\text{QWP}$, $0^\circ_\text{HWP}$, $0^\circ_\text{QWP}$ \\[4pt]

(2) $0^\circ_\text{QWP}$, $22.5^\circ_\text{HWP}$
  & (2) $22.5^\circ_\text{HWP}$
  & (2) $22.5^\circ_\text{HWP}$
  & (2) $0^\circ_\text{QWP}$, $0^\circ_\text{HWP}$, $45^\circ_\text{QWP}$ \\[4pt]

(3) $45^\circ_\text{QWP}$, $0^\circ_\text{HWP}$
  &
  & (3) $45^\circ_\text{HWP}$
  & (3) $0^\circ_\text{QWP}$, $45^\circ_\text{HWP}$, $0^\circ_\text{QWP}$ \\[4pt]

  &   &
  & (4) $45^\circ_\text{QWP}$, $0^\circ_\text{HWP}$, $0^\circ_\text{QWP}$ \\[4pt]

  &   &
  & (5) $45^\circ_\text{QWP}$, $0^\circ_\text{HWP}$, $90^\circ_\text{QWP}$ \\[4pt]

  &   &
  & (6) $45^\circ_\text{QWP}$, $45^\circ_\text{HWP}$, $90^\circ_\text{QWP}$ \\[4pt]

  &   &
  & (7) $90^\circ_\text{QWP}$, $0^\circ_\text{HWP}$, $0^\circ_\text{QWP}$ \\[4pt]

  &   &
  & (8) $90^\circ_\text{QWP}$, $0^\circ_\text{HWP}$, $45^\circ_\text{QWP}$ \\[4pt]

  &   &
  & (9) $90^\circ_\text{QWP}$, $45^\circ_\text{HWP}$, $0^\circ_\text{QWP}$ \\[4pt]

  &   &
  & (10) $90^\circ_\text{QWP}$, $45^\circ_\text{HWP}$, $45^\circ_\text{QWP}$ \\[4pt]
  
\hline
\end{tabular}
\end{table*}

\newpage

\end{document}